\documentclass[12pt]{article}
\usepackage{amsfonts}
\usepackage{amsmath,amssymb}

\def\sfrac#1#2{{\textstyle\frac{#1}{#2}}}
\def\dfrac#1#2{{\displaystyle\frac{#1}{#2}}}

\begin{document}
\numberwithin{equation}{section}

\title{Surface waves \\ in a deformed isotropic hyperelastic material \\
subject to an isotropic internal  constraint}
\author{Michel Destrade,  Nigel H. Scott}
\date{2004}
\maketitle

\bigskip

\begin{abstract}
An isotropic elastic half space is prestrained so that two of the 
principal axes of strain lie in the bounding plane, which itself 
remains free of traction.  
The material is subject to an isotropic constraint of arbitrary 
nature.  
A surface wave is propagated sinusoidally along the bounding surface 
in the direction of a principal axis of strain and decays away from 
the surface.  
The exact secular equation is derived by a direct method  for such a 
principal surface wave; 
it is cubic in a quantity whose square is linearly related to the 
squared wave speed.   
For the prestrained material,  replacing the squared wave speed by 
zero  gives an explicit bifurcation, or stability, criterion.  
Conditions on the existence and uniqueness of surface waves are 
given.  
The bifurcation criterion is derived for specific strain energies 
in the case of  four isotropic constraints: 
those of incompressibility, Bell, constant area, and Ericksen.  
In each case investigated, the bifurcation criterion is found to be 
of a universal nature in that it depends only on the principal 
stretches, not on the material constants.
Some results related to the surface stability of arterial wall 
mechanics are also presented.

\end{abstract}

\newpage

%
\section{Introduction}
\setcounter{equation}{0}
  
There has been a large body of literature on surface waves 
propagating sinusoidally  along the bounding planar surface of an 
elastic half space with attenuation of the wave amplitude in the 
direction normal to the bounding plane.  
The earliest paper is that of Rayleigh \cite{Rayl85}, who 
investigated surface waves propagating across the surface of an 
isotropic earth in the context of seismology.  
The modern theory of surface waves derives in large measure from the 
sextic formalism of Stroh \cite{Stroh01}.  
This approach has been employed by many different authors to address 
many different problems of linear anisotropic elasticity 
and the results have been comprehensively reviewed by Ting 
\cite{Ting}.  
In parallel, tremendous progress has been made in the theory of 
small-amplitude surface waves propagating on finitely deformed, 
nonlinearly elastic half-spaces, from the seminal works of Hayes and 
Rivlin \cite{HaRi61} (compressible materials) to those of Dowaikh and 
Ogden \cite{DoOg90} and many others.

In the present paper we consider an isotropic elastic half space 
prestrained so that two of the principal axes of strain lie in the 
bounding plane which itself remains free of traction.  
Additionally, the material is subject to an isotropic constraint of 
an \textit{arbitrary} nature.  
One example of such a constraint is incompressibility, 
which has received much attention in the 
literature, and another is the Bell constraint, whose experimental and 
theoretical  properties have been thoroughly reviewed by Beatty 
\cite[Chap. 2]{Beat01}.  
A wave is propagated sinusoidally in the direction of one of the 
principal axes lying in the bounding plane, has no in-plane 
displacement component in the direction of the second principal axis 
lying in this plane, but has attenuating amplitude in the direction 
of the third principal axis, orthogonal to this plane 
(principal surface wave).  
The theoretical framework for the study of these waves is set up in 
Section 2, including a discussion of four example of isotropic 
constraints, the incremental equations of motion and of constraint, 
and the strong ellipticity condition.  

In Section 3 we derive an explicit secular equation for these 
principal surface waves that is cubic in a quantity whose square is 
linearly related to the squared wave speed.  
On restricting attention to an unstrained isotropic material we find 
that the secular equation reduces to that found by Rayleigh 
\cite{Rayl85} in the incompressible case.  
We show that for the unstrained material all isotropic constraints 
are the same, that is all reduce to incompressibility \cite{PoVi89}.  
Returning to the general secular equation of the prestrained material 
and replacing the squared wave speed by zero we obtain an explicit 
bifurcation, or stability, criterion for the material.  
Some results on the existence and uniqueness of surface waves are 
given.

Finally, in Section 4 we examine the bifurcation criterion for each 
of four examples of isotropic constraint and obtain explicit results 
by choosing special forms of the strain energy function.  
In all cases considered we find that the bifurcation criterion is of 
a universal nature in that it depends only on the principal 
stretches, not on the material constants.

\section{Preliminaries}
\setcounter{equation}{0}
\label{Preliminaries}

\subsection{Deformed constrained half-space with a free plane surface}
\label{Hyperelastic materials}

We consider a semi-infinite body made of homogeneous isotropic 
hyperelastic material at rest in a configuration $\mathcal{B}_u$ with 
strain energy density $W$ per unit volume of $\mathcal{B}_u$ and mass 
density $\rho$.
Let $(O, X_1, X_2, X_3)$ 
be a fixed rectangular Cartesian coordinate system such that the body
occupies the region $X_2 \ge 0$.  
The orthonormal set of vectors $\{{\bf i}, {\bf j}, {\bf k} \}$ is 
aligned with the coordinate axes.

Loads $P_1$, $P_2$, $P_3$ are applied at infinity to 
deform and maintain the half-space in a static state $\mathcal{B}_e$
of finite pure homogeneous deformation, with corresponding stretch
ratios $\lambda_1$,  $\lambda_2$,  $\lambda_3$ in the  
$\bf{i}, \bf{j}, \bf{k}$ directions.
Thus, the position of a particle at $(X_1, X_2, X_3)$ in 
$\mathcal{B}_u$ is at 
$(\overline{x}_1, \overline{x}_2, \overline{x}_3)$ 
in $\mathcal{B}_e$, where
$\overline{x}_1 = \lambda_1 X_1$, 
$\overline{x}_2 = \lambda_2 X_2$,  
$\overline{x}_3 = \lambda_3 X_3$.
The constant deformation gradient associated with the deformation is
\begin{equation}
\overline{\bf{F}}=
 \lambda_1 \bf{i}\otimes\bf{i} + 
  \lambda_2 \bf{j}\otimes\bf{j} + 
   \lambda_3 \bf{k}\otimes\bf{k}.
\end{equation}

In an isotropic hyperelastic material the strain energy is a symmetric 
function $W(\lambda_1,\lambda_2,\lambda_3)$ of the principal 
stretches, i.e. its value is left unchanged by any permutation of the 
stretches $\lambda_1, \lambda_2, \lambda_3$.
The material is subject to an isotropic internal constraint, 
written as
\begin{equation}\label{Constraint}
\Gamma (\lambda_1,  \lambda_2,  \lambda_3) = 0,
\end{equation}
in which $\Gamma$ is a symmetric function of the principal stretches 
$\lambda_i$.
We restrict attention to  constraints such that:
\begin{equation}
\Gamma_i >0, \;\;\; {\rm where} \;\;\; 
\Gamma_i := \partial\Gamma/\partial\lambda_i.
\end{equation} 

Four examples of such constraints are treated explicitly in this paper
and they are henceforward denoted by roman numerals:
the incompressibility (I), Bell (II), areal (III), and Ericksen (IV)
constraints,
\begin{align} \label{ConstraintsExamples}
&   \Gamma^I  :=  \lambda_1   \lambda_2   \lambda_3 - 1 = 0, 
&&  \Gamma^{II}  :=  \lambda_1 +  \lambda_2 +  \lambda_3 - 3 = 0, 
\nonumber \\
&   \Gamma^{III} :=  \lambda_1 \lambda_2 +  \lambda_2 \lambda_3 
                       +  \lambda_3\lambda_1  - 3 = 0,
&&  \Gamma^{IV}  :=  \lambda_1^2 +  \lambda_2^2 +  \lambda_3^2 - 3 = 0.
\end{align}
 The incompressibility constraint is often used for the modelling of
finite deformations of rubber-like materials and shows good correlation
with experiment (see for instance Ogden \cite[Chap. 7]{Ogde84}).  
The Bell constraint was found to hold experimentally over countless  
trials on polycrystalline annealed solids, including aluminum, brass, 
copper, and mild steel, see Beatty \cite[Chap. 2]{Beat01}.
The areal (or constant area) constraint has the interpretation that a 
material cube in the reference configuration with edges parallel to the principal axes of strain  retains the same total surface area  after deformation; 
it was studied from a purely mathematical point of
view by Bosi and Salvatori \cite{BoSa96}.  
Finally, the fourth constraint was proposed by Ericksen \cite{Eric86}
to model the behaviour of  certain twinned elastic crystals.
Although Ericksen proposed a multi-constrained model,  
pursued by Scott \cite{Scot92} in the context of wave propagation, 
some authors \cite{Antm95,Sacc01} refer to  the single constraint 
(\ref{ConstraintsExamples})$_4$ in nonlinear elasticity theory
as `Ericksen's constraint'.

The general constraint (\ref{Constraint}) generates the workless
reaction tensor $\overline{\bf{N}}$, see \cite{ChWB85}, given by
$\overline{\bf{N}} = J^{-1} \overline{\bf{F}}
  (\partial \Gamma/\partial \overline{\bf{F}})^{\rm {T}}$,
where $J = \lambda_1 \lambda_2 \lambda_3$.
Explicitly, the non-zero components of $\overline{\bf{N}}$ are
\begin{equation}\label{2.6}
\overline{N}_{ii} = J^{-1} \lambda_i \Gamma_i \;\;\; \mbox{(no sum)}.
\end{equation} 
Thus for the four examples of constraints (\ref{ConstraintsExamples})
we find the following constraint tensors:
\begin{equation}  \label{ConstraintsTensorExamples}
\overline{\mathbf{N}}^I = \mathbf{1}, 
\quad 
\overline{\mathbf{N}}^{II} =  J^{-1}\mathbf{V}, 
\quad 
\overline{\mathbf{N}}^{III}  
    = J^{-1}[(\text{tr } \mathbf{V}) \mathbf{V} - \mathbf{V}^2],
\quad 
\overline{\mathbf{N}}^{IV} = 2J^{-1}\mathbf{V}^2,
\end{equation}
in terms of the left stretch tensor 
$\bf{V} = {\rm diag}\,(\lambda_1, \lambda_2, \lambda_3)$.

Associated with the deformation 
$\mathcal{B}_u \rightarrow \mathcal{B}_e$ is the Cauchy stress tensor
$\overline{\mbox{\boldmath $\sigma$}}$ which takes diagonal form with 
non-zero components 
\begin{equation}
\overline{\sigma}_{ii}=J^{-1} \lambda_i W_i 
   + \overline{P}\,\overline{N}_{ii}
 \;\;\; \mbox{(no sum)},
\end{equation}
where $ W_i := \partial W/\partial\lambda_i$ and $\overline{P}$ is a 
scalar to be determined from the equations of equilibrium and boundary 
conditions as follows.
First, we note that for $\overline{P}$ constant, the equations of 
equilibrium 
$\partial \overline{\sigma}_{ij}/\partial \overline{x}_j = 0$ are 
automatically satisfied.
Next, we assume that the surface $\overline{x}_2=0$ is free of 
tractions; it follows that $\overline{\sigma}_{22}=0$ (and $P_2=0$) 
and so 
\begin{equation} \label{Pbar}
\overline{P} = - J^{-1}\lambda_2W_2/\overline{N}_{22} = -W_2/\Gamma_2.
\end{equation}
Consequently, the constant loads $P_1$, 
$P_3$ needed at infinity in
order to maintain the half-space in the deformed configuration 
$\mathcal{B}_e$ are ($k=1,3$ no sum),
\begin{equation}
P_k  = -\overline{\sigma}_{kk} 
  = \left(W_2/\Gamma_2\right)\overline{N}_{kk} - J^{-1} \lambda_k W_k
  = \left(W_2 \Gamma_k - W_k \Gamma_2 \right)\lambda_k/ (J \Gamma_2).
\end{equation}
Note that when the boundary surface is not free of tractions 
($\overline{\sigma}_{22} \ne 0$) then $\overline{P}$ in \eqref{Pbar} 
must be replaced by $\overline{P}
    = (J\overline{\sigma}_{22} - \lambda_2 W_2)/(\lambda_2 \Gamma_2)$. 
However, the primary interest of this paper is in studying the 
influence of internal constraints upon the propagation of surface  
waves rather than the influence of pre-loading, and we assume 
henceforward that \eqref{Pbar} holds.

\subsection{Superposed infinitesimal motion and strong 
ellipticity condition}

We now consider the propagation of an incremental motion in the 
deformed half-space $\mathcal{B}_e \rightarrow \mathcal{B}_t$,
described by
\begin{equation}
\bf{x} = \overline{\bf{x}}
 + \epsilon \bf{u}(\overline{\bf{x}},t),
\end{equation}
where $\bf{x}$ is the position in $\mathcal{B}_t$ of a particle 
which was at $ \overline{\bf{x}}$ in $\mathcal{B}_e$ and $\bf{u}$ is 
referred to as  the displacement vector.
The parameter $\epsilon$ is small, so that terms of order higher than
one in $\epsilon$ may be neglected.
The linear fourth-order instantaneous elasticity tensor $\bf{B}^*$ 
associated with the motion is \cite[(2.19)]{ChWB85}
\begin{equation} \label{B^*}
B^*_{ijkl} = B_{ijkl} + \overline{P} \check{B}_{ijkl}
           = B_{ijkl} - (W_2/\Gamma_2) \check{B}_{ijkl},
\end{equation}
where the non-zero components of $\bf{B}$ and $\bf\check{B}$ are given 
by \cite[(6.3.15)]{Ogde84}
\begin{equation}\label{B*}
\begin{array}{lll}
JB_{iijj} = \lambda_i \lambda_j W_{ij}, 
&\;\;\; &J\check{B}_{iijj}  = \lambda_i \lambda_j \Gamma_{ij},  \\
JB_{ijij} = 
    \dfrac{\lambda_i W_i-\lambda_j W_j}
        {\lambda_i^2-\lambda_j^2}\lambda_i^2,
&\;\;\;& J\check{B}_{ijij} = 
    \dfrac{\lambda_i \Gamma_i-\lambda_j \Gamma_j}
        {\lambda_i^2-\lambda_j^2}\lambda_i^2, \;\;\;(i\neq j)\\
JB_{ijji} = JB_{ijij} - \lambda_i W_i,
&\;\;\;& J\check{B}_{ijji} = J\check{B}_{ijij} -\lambda_i \Gamma_i,
\;\;\; (i\neq j) 
\end{array}
\end{equation}
and there is no summation over $i$ or $j$.  Here the second line  
holds when $i \ne j$, 
$\lambda_i \ne \lambda_j$; in the case where $i \ne j$, 
$\lambda_i = \lambda_j$, it must be replaced by
$JB_{ijij} =\frac12(JB_{iiii} - JB_{iijj} + \lambda_i W_i)$, 
$J\check{B}_{ijij}  =\frac12 (J\check{B}_{iiii} 
- J\check{B}_{iijj} + \lambda_i \Gamma_i)$, 
see \cite[(6.3.16)]{Ogde84}.
Because of hyperelasticity the symmetries $B^*_{ijkl} = B^*_{klij}$ 
hold so that in particular we have
\begin{equation}
B^*_{iijj} = B^*_{jjii},\;\;\;
B^*_{ijji} = B^*_{jiij}\;\;\;\mbox{(no sum)}.
\label{sym}
\end{equation}

The incremental nominal stress associated with this motion is 
$\bf{s}$ given by \cite[(3.10)]{ChWB85}
\begin{equation} \label{nominalStress}
s_{ij} = \overline{\sigma}_{ij} + B^*_{ijkl}u_{l,k}
            + p \overline{N}_{ij},
\end{equation}
where $p$ represents the increment in $\overline{P}$.

The equations of motion, 
together with the incremental constraint, read, from  
\cite[(3.13)]{ChWB85},
\begin{equation} \label{motion1}
\overline{\rho} u_{j,tt} = s_{ij,i}, \;\;\;
\overline{N}_{11} u_{1,1}
 + \overline{N}_{22} u_{2,2}
  + \overline{N}_{33} u_{3,3} =0. 
\end{equation}
 
For future convenience, we assume that $\bf{B}^*$ is strongly
elliptic, that  is, 
\begin{equation} \label{SEgeneral}
B^*_{ijkl} m_i n_j m_k n_l > 0,
\;\;\; \mbox{for all} \;\;\; {\bf m}, {\bf n}
\;\;\; \mbox{such that} \;\;\;
m_i\overline{N}_{ij}n_j =0.
\end{equation}
The vectors
\[
{\bf m}
  = \overline{N}_{11}^{- {\frac{1}{2}}} \cos\theta \,{\bf i} 
  +\overline{N}_{22}^{- {\frac{1}{2}}} \sin \theta\, {\bf j}
= J^{ {\frac{1}{2}}}
        [(\lambda_1\Gamma_1)^{- {\frac{1}{2}}}
     \cos\theta\, {\bf i} 
  + (\lambda_2\Gamma_2)^{- {\frac{1}{2}}}
                               \sin \theta\, {\bf j}],
\]
\begin{equation}
{\bf n}
 = -\overline{N}_{11}^{- {\frac{1}{2}}} \sin\theta \,{\bf i}
 + \overline{N}_{22}^{- {\frac{1}{2}}} \cos \theta \,{\bf j}
= J^{ {\frac{1}{2}}}
        [-(\lambda_1\Gamma_1)^{- {\frac{1}{2}}}
                                        \sin\theta \,{\bf i} 
  + (\lambda_2\Gamma_2)^{- {\frac{1}{2}}} 
                             \cos \theta \,{\bf j}],
\end{equation}
$0 \le \theta \le 2 \pi$,
are two vectors satisfying (\ref{SEgeneral})$_2$.
Introducing the quantities $A$, $B$, $C$, defined by
\begin{equation}  \label{ABC}
A = (\lambda_1\lambda_2\Gamma_1\Gamma_2)^{-1} B^*_{1212}, \;\;\;
C = (\lambda_1\lambda_2\Gamma_1\Gamma_2)^{-1} B^*_{2121}, 
\end{equation}
\[
B 
 = {\sfrac{1}{2}}
    [(\lambda_1\Gamma_1)^{-2} B^*_{1111}
         + (\lambda_2\Gamma_2)^{-2} B^*_{2222}]
   -(\lambda_1\lambda_2\Gamma_1\Gamma_2)^{-1}
                              (B^*_{1122} + B^*_{1221}),
\]
we obtain from (\ref{SEgeneral}) the inequality
\begin{equation}
A \cos^4 \theta + 2B \sin^2 \theta \cos^2 \theta + C \sin^4 \theta >0,
\end{equation}
holding for all $\theta$. 
In particular, the choices $\theta = 0$, $\pi/2$, 
$\arctan (-B/C)^{- 1/2}$, give in turn
\begin{equation} \label{SE1}
A >0, \;\;\; C>0, \;\;\;  B + \sqrt{AC}>0.
\end{equation}
Using different notations, similar inequalities are given by
Knowles and Sternberg \cite{KnSt77} for unconstrained materials,
by Ogden \cite{Ogde84} for incompressible materials,
and by Destrade \cite{MD2} for Bell materials.

\section{Surface waves}
\label{SurfaceWaves}
\setcounter{equation}{0}


Here we specialize the equations of motion to the consideration of an 
inhomogeneous principal plane wave traveling over the free surface 
$\overline{x}_2=0$ and derive 
the corresponding secular equation.

\subsection{Equations of motion and boundary conditions}

A surface (Rayleigh) wave travels sinusoidally with time $t$ in a 
direction parallel to the plane $\overline{x}_2=0$  ($X_2=0$) leaving 
it free of tractions  and decaying away 
from this surface as $\overline{x}_2\rightarrow\infty$.
For simplicity, we consider a wave propagating  with speed $v$ and 
wave number $k$ in the principal direction of prestrain 
$OX_1 = O\overline{x}_1$.  We refer to this as a principal wave. 
For this wave, antiplane strain decouples from inplane 
strain, and the displacement components are of the form:
\begin{equation}
u_i = U_i(k\overline{x}_2)e^{ik(\overline{x}_1 - vt)} \;\;\; (i=1,2), 
\;\;\;
u_3=0, \;\;\; 
p= kQ(k\overline{x}_2)e^{ik(\overline{x}_1-vt)},
\end{equation}
in which $U_1$, $U_2$ and $Q$ are functions of $k\overline{x}_2$ to be 
determined.
Similarly, the  antiplane stress decouples from the plane stress,
and the equations of motion (\ref{motion1}) reduce to 
\begin{equation} \label{motion2}
 s_{11,1} + s_{21,2} = \overline{\rho} u_{1,tt}, \;\;\;
 s_{12,1} + s_{22,2} = \overline{\rho} u_{2,tt}, \;\;\;
\overline{N}_{11} u_{1,1} + \overline{N}_{22} u_{2,2} =0. 
\end{equation}

Because the $\overline{\sigma}_{ij}$ terms in the components 
(\ref{nominalStress}) of $s_{ij}$ are constant, we write the 
relevant stress components in the form
\begin{equation}\label{3.3}
s_{ij} = \overline{\sigma}_{ij}
 + kS_{ij}(k\overline{x}_2)e^{ik(\overline{x}_1-vt)} \;\;\; (i,j=1,2),
\end{equation}
and the equations of motion reduce further to 
 \begin{equation} \label{motion3}
 iS_{11} + S'_{21} = -\overline{\rho} v^2 U_1, \;\;\;
 iS_{12} + S'_{22} = -\overline{\rho} v^2 U_2, \;\;\;
i \lambda_1 \Gamma_1 U_1 + \lambda_2 \Gamma_2 U'_2 =0. 
\end{equation}
Explicitly, the $S_{ij}$ are given from (\ref{nominalStress}) and 
(\ref{3.3})  by
\begin{align} \label{3.5}
&  S_{11} = i B^*_{1111} U_1 + B^*_{1122} U'_2 + Q \overline{N}_{11},
&& S_{12} = B^*_{1221} U'_1 + iB^*_{1212} U_2,
\nonumber \\
&  S_{22} = i B^*_{1122} U_1 + B^*_{2222} U'_2 + Q \overline{N}_{22},
&& S_{21} = B^*_{2121} U'_1 + iB^*_{1221} U_2,
\end{align}
in which the symmetries (\ref{sym}) have been used.

Now we use  (\ref{motion3}) and (\ref{3.5}) to formulate the problem
as a system of four first-order differential equations for the 
unknown functions $U_1$, $U_2$, $S_{21}$, $S_{22}$:
\begin{align}\label{motion4}
& U_1'  = -i \dfrac{B^*_{1221}}{B^*_{2121}}U_2 + 
             \dfrac{1}{B^*_{2121}}S_{21},
\quad 
U'_2 = -i\dfrac{\lambda_1\Gamma_1}{\lambda_2\Gamma_2}U_1,
\nonumber \\
& S_{21}'
 =\left[B^*_{1111} 
 - 2\dfrac{\lambda_1\Gamma_1}{\lambda_2\Gamma_2}B^*_{1122}
     +  \left(
       \dfrac{\lambda_1\Gamma_1}{\lambda_2\Gamma_2}\right)^2 B^*_{2222}
                  -\overline{\rho} v^2\right]U_1 
            - i\dfrac{\lambda_1\Gamma_1}{\lambda_2\Gamma_2}S_{22},
\nonumber \\
& S_{22}'
 = \left[\dfrac{B^*_{2121}B^*_{1212} - B^{*2}_{1221}}{B^*_{2121}}
                                      -\overline{\rho} v^2\right]U_2 
            - i\dfrac{B^*_{1221}}{B^*_{2121}}S_{21}.
\end{align}
Equations (\ref{motion4})$_1$ and (\ref{motion4})$_2$ are obtained 
from (\ref{3.5})$_4$ and (\ref{motion3})$_3$, respectively, and 
(\ref{2.6}) is employed in the latter.  
Equation (\ref{motion4})$_3$ is obtained by eliminating $Q$ between 
(\ref{3.5})$_1$ and  (\ref{3.5})$_3$ using  (\ref{2.6}), and then 
eliminating $S_{11}$ between this equation and (\ref{motion3})$_1$.  
Finally, (\ref{motion4})$_4$ is obtained by eliminating $S_{12}$ 
between (\ref{motion3})$_2$ and (\ref{3.5})$_2$ and then using 
(\ref{3.5})$_4$ to eliminate $U'_1$ in favor of $S_{21}$. 

Equations (\ref{motion4}) are subject to the boundary conditions of 
decay as $\overline{x}_2 \rightarrow \infty$ and of vanishing  
traction on $\overline{x}_2=0$:
\begin{equation} \label{BC}
S_{21}(0)=S_{22}(0)=0.
\end{equation}

Furthermore, each unknown may be written in terms of  a single unknown
function $\varphi$ of the variable 
\[
z = \frac{ \lambda_1\Gamma_1}{\lambda_2\Gamma_2}k\overline{x}_2,
\]
with prime now denoting differentiation with respect to $z$:
\begin{align} \label{U-S-phi}
& U_1 =i\varphi'(z),
 \quad
U_2 = \varphi(z),
 \quad
S_{21} 
 =i \frac{\lambda_1\Gamma_1}{\lambda_2\Gamma_2}B^*_{2121}\varphi''(z)
  + i B^*_{1221} \varphi(z),
 \\
& S_{22} = 
- \frac{\lambda_1\Gamma_1}{\lambda_2\Gamma_2} B^*_{2121}\varphi'''(z) 
\nonumber \\
& \;\;\;\;\;\;\;\;\;\;\mbox{}
 + \frac{\lambda_2\Gamma_2}{\lambda_1\Gamma_1}
    \left[B^*_{1111} 
 -\frac{\lambda_1\Gamma_1}{\lambda_2\Gamma_2}(B^*_{1221} + 2B^*_{1122})
 +\left(
      \frac{\lambda_1\Gamma_1}{\lambda_2\Gamma_2}\right)^2 B^*_{2222}
                               -\overline{\rho} v^2\right] \varphi'(z).
\nonumber
\end{align}
Here, the expressions for $U_1$, $S_{21}$, and $S_{22}$ were obtained 
from (\ref{motion4})$_{2,1,3}$, respectively.
The last equation of this set, namely (\ref{motion4})$_4$, then yields 
a differential equation for $\varphi(z)$:
\begin{equation} \label{motion5}
\gamma^* \varphi'''' - (2\beta^*-\overline{\rho} v^2) \varphi''
  + (\alpha^*-\overline{\rho} v^2) \varphi =0,
\end{equation}
where
\begin{align} \label{AlphaBetaGamma}
& \alpha^* = B^*_{1212}, 
\quad
\gamma^* = 
    \left(
   \frac{\lambda_1\Gamma_1}{\lambda_2\Gamma_2}\right)^2 B^*_{2121} 
        =  \frac{\Gamma_1^2}{\Gamma_2^2} \alpha^*,
\nonumber \\
& 2\beta^* = B^*_{1111} 
 -2\frac{\lambda_1\Gamma_1}{\lambda_2\Gamma_2}(B^*_{1221} + B^*_{1122})
     +\left(
   \frac{\lambda_1\Gamma_1}{\lambda_2\Gamma_2}\right)^2 B^*_{2222}.
\end{align}
We have used the property $
\lambda^2_iB^*_{jiji} = \lambda^2_jB^*_{ijij}\;
(i\neq j,\;\mbox{no sum})$ derived from (\ref{B*}).
By comparison with the quantities $A$, $B$, $C$, defined in 
(\ref{ABC}), and use of the consequences (\ref{SE1}) of the strong 
ellipticity (S-E) condition (\ref{SEgeneral}), we find that these 
coefficients satisfy the inequalities
\begin{equation} \label{SE2}
\alpha^*>0, \;\;\; \gamma^*>0, 
      \;\;\; \beta^* + \sqrt{\alpha^*\gamma^*}>0.
\end{equation}

\subsection{Secular equation and bifurcation criterion}

We now derive the exact form of the secular equation for a surface 
wave traveling in a principal direction of  a 
deformed isotropic material subject to a single isotropic constraint.
Because the wave amplitude  decays as $z\rightarrow\infty$  away from 
the free plane $z = 0$, we seek a solution for $\varphi$ in the form 
\begin{equation}\label{3.12}
\varphi(z) = A_1 e^{-s_1z} + A_2 e^{-s_2z}, \;\;\; \Re (s_i)>0, 
\;\;\; s_1 \ne s_2.
\end{equation}
From  (\ref{motion5}), the $s_i$ are roots of the biquadratic
\begin{equation} \label{quartic}
\gamma^* s^4 - (2\beta^*-\overline{\rho} v^2)s^2
  + (\alpha^*-\overline{\rho} v^2) =0.
\end{equation}
The roots $s_i^2$ of this real quadratic are either both real 
(and, if so, both positive  because we must have $\Re (s_i)>0$) or 
they are a complex conjugate pair.   
In either case, $s_1^2  s_2^2>0$ and so, by (\ref{quartic}),
\begin{equation}
0 \le \overline{\rho} v^2 \le \alpha^*.
\end{equation}
Although necessary, this inequality on the squared wave speed  is 
not sufficient to ensure the decay of the wave amplitude, as is seen 
in the next subsection.

The boundary conditions (\ref{BC}) yield, using 
(\ref{U-S-phi})$_{3,4}$, (\ref{AlphaBetaGamma}), and 
(\ref{quartic}),
\begin{align} \label{3.15}
&\left(\gamma^* s_1^2
     + \frac{\lambda_1\Gamma_1}{\lambda_2\Gamma_2}B^*_{1221}\right)A_1
+ \left(\gamma^* s_2^2
    + \frac{\lambda_1\Gamma_1}{\lambda_2\Gamma_2}B^*_{1221}\right)A_2
=0, 
\nonumber\\
& s_1\left(\gamma^* s_2^2 
    + \frac{\lambda_1\Gamma_1}{\lambda_2\Gamma_2}B^*_{1221}\right)A_1
+  s_2\left(\gamma^* s_1^2 
    + \frac{\lambda_1\Gamma_1}{\lambda_2\Gamma_2}B^*_{1221}\right)A_2
  =0.
\end{align}
The vanishing of the determinant of this homogeneous system  gives 
the {\em secular equation}, that is, the equation for the wave speed.
Introducing the quantities
\begin{equation} \label{EtaDelta}
\eta^* = \sqrt{\frac{\alpha^*-\overline{\rho} v^2}{\gamma^*}},
\quad
\delta^* = \frac{\lambda_1\Gamma_1}{\lambda_2\Gamma_2}B^*_{1221}
 = \frac{\lambda_2\Gamma_2}{\lambda_1\Gamma_1}\gamma^*,
\end{equation}
and using (\ref{quartic}) we obtain from the vanishing of the 
determinant of (\ref{3.15}), 
\begin{equation}\label{3.17}
f(\eta^*) := \eta^{*3} + \eta^{*2}
  + \frac{2\beta^*+2\delta^* - \alpha^*}{\gamma^*}\eta^*
     -\frac{\delta^{*2}}{\gamma^{*2}} = 0,
\end{equation}
after removing the $s_1-s_2$ factor.  
Equation (\ref{3.17}) is the required secular equation.

In the special case where the biquadratic (\ref{quartic}) has double 
roots, so that $s_2^2=s_1^2$, the above derivation is no longer valid.
The form (\ref{3.12}) of solution   must be replaced by
\[
\varphi(z)=(A_1+A_2z)e^{-s_1z}
\]
but it can still be shown that the secular equation is given by 
(\ref{3.17}).  
Thus (\ref{3.17}) furnishes the secular equation in all cases.

The cubic equation (\ref{3.17}) in $\eta^*$ has remarkable features: 
the coefficients of the two highest powers are each always equal to 
unity, irrespective of the pre-deformation, constraint, and strain 
energy function. 
By \eqref{EtaDelta}$_3$, the term independent of $\eta^*$ depends 
only on $\lambda_1$, $\lambda_2$, and the constraint, but \textit{not} 
on the strain energy function. 
For instance, for the I-IV constraints \eqref{ConstraintsExamples}, 
we find that 
\begin{equation} 
\frac{\delta^*}{\gamma^*} = 1, 
\quad 
\frac{\lambda_2}{\lambda_1}, 
\quad 
\frac{\lambda_1 + \lambda_3}{\lambda_2 + \lambda_3} 
    \frac{\lambda_2}{\lambda_1},
\quad 
\frac{\lambda_2^2}{\lambda_1^2},
\end{equation}
respectively, whatever $W$ may be.  
From (\ref{AlphaBetaGamma})$_2$ it is clear also that 
$\alpha^*/\gamma^*$ is independent of $W$ and it follows that the 
secular equation (\ref{3.17}) depends on $W$ only through the term 
$\beta^*/\gamma^*$ appearing in the coefficient of the term linear in 
$\eta^*$.

In an undeformed material ($\lambda_1=\lambda_2=\lambda_3=1$), 
we have $\alpha^*=\beta^*=\gamma^*=\delta^*=\mu$, where $\mu$ is the 
infinitesimal shear modulus, and the secular equation (\ref{3.17}) 
reduces to 
\begin{equation}\label{3.18}
f(\eta) := \eta^3 + \eta^2 + 3\eta - 1 = 0, 
\quad \eta = \sqrt{1 - \overline{\rho}v^2/\mu}.
\end{equation}
The unique positive real root of  this equation corresponds to 
$\overline{\rho}v^2/\mu \approx  0.9126$, 
in accordance with Rayleigh's result \cite{Rayl85} for incompressible
linear isotropic elastic materials.

Equation (\ref{3.18}) is valid for all isotropic constraints provided 
that the isotropic elastic material is undeformed.  
In fact, in the undeformed state we can show that all isotropic 
constraints are equivalent by arguing as follows (see Podio-Guidugli 
and Vianello \cite{PoVi89} for an alternative treatment).  
The constraint is $\Gamma(\lambda_1, \lambda_2, \lambda_3) = 0$ where 
the function  $\Gamma$ is symmetric in its arguments, i.e. its value 
is left unchanged by any permutation of the stretches 
$\lambda_1, \lambda_2, \lambda_3$.  
For small strains write $\lambda_i =1 +  e_{i}$ where 
$e_{i}$ is  the extension ratio.  
The constraint holds for $\lambda_i = 1$ and for 
$\lambda_i = 1 + e_i$ (for each $i$) so an application of Taylor's 
theorem gives
\[ 
\frac{\partial\Gamma}{\partial\lambda_1}e_1 
 + \frac{\partial\Gamma}{\partial\lambda_2}e_2 
  + \frac{\partial\Gamma}{\partial\lambda_3}e_3 = 0,
\]
where terms quadratic in $e_i$ are neglected.   
The partial derivatives are evaluated at $\lambda_i = 1$  and, 
because of the symmetry condition on $\Gamma$, are all equal.  
Thus, in the undeformed state, each isotropic constraint takes the 
form
\begin{equation}
e_1+e_2+e_3 = 0
\label{3.18a}
\end{equation}
of the constraint of incompressibility for infinitesimal deformations.

The {\em bifurcation criterion} is obtained by writing $v=0$ in the 
secular equation  (\ref{3.17}) and indicates when the half-space might 
become unstable \cite{Biot65}:
\begin{equation} \label{secular}
f\left(\sqrt{\frac{\alpha^*}{\gamma^*}}\right) = 
\left(\frac{\alpha^*}{\gamma^*}\right)^{\frac{3}{2}} 
+ \frac{\alpha^*}{\gamma^*}
+\frac{2\beta^*+2\delta^*-\alpha^*}{\gamma^*}
    \sqrt{\frac{\alpha^*}{\gamma^*}}
     -\frac{\delta^{*2}}{\gamma^{*2}} = 0,
\end{equation}
which becomes 
\begin{equation}\label{3.20}
\frac{\Gamma_2^2}{\Gamma_1^2}
+ 2\frac{\beta^*+\delta^*}{\gamma^*}\frac{\Gamma_2}{\Gamma_1}
     -\frac{\delta^{*2}}{\gamma^{*2}} = 0
\end{equation}
on using (\ref{AlphaBetaGamma})$_2$ to eliminate $\alpha^*/\gamma^*$.  
Using  (\ref{B^*})$_2$, (\ref{B*}), (\ref{AlphaBetaGamma}) and 
(\ref{EtaDelta})$_2$,   we may rewrite (\ref{3.20}) in terms of the 
derivatives with respect to $\lambda_i$ of the strain energy function 
$W$ and  the constraint $\Gamma$:
\begin{multline} \label{bifurcation}
\Gamma_2^2 W_{11} - 2\Gamma_1\Gamma_2 W_{12}
  + \Gamma_1^2 W_{22} 
\\
+ \Gamma_1(\Gamma_2 W_1 -\Gamma_1 W_2)/\lambda_1
  - (\Gamma_{11} \Gamma_2^2 -2\Gamma_1\Gamma_2\Gamma_{12}
      +\Gamma_{22} \Gamma_1^2)  W_2 /\Gamma_2  = 0,
\end{multline}
an explicit form of the bifurcation criterion.

\subsection{Existence and uniqueness of a surface wave}

Following Chadwick \cite{Chad95} we recast our secular equation 
\eqref{3.17} as 
\begin{multline}
f(v) =  \text{det }\begin{bmatrix}
                   \eta^* r^* & \dfrac{\delta^*}{\gamma^*} - \eta^* \\
                   \dfrac{\delta^*}{\gamma^*} - \eta^* &  r^*
                   \end{bmatrix}  
= 0, \\ \text{ where }  \qquad 
r^* = 
 \sqrt{(\eta^*+1)^2-\frac{\alpha^*+ \gamma^* - 2\beta^*}{\gamma^*}}.
\end{multline}
Chadwick used this ``matrix reformulation'' of the secular equation 
in the case of incompressible (I) materials (where  
$\delta^* / \gamma^* = 1$) to derive, in a rigorous manner, essential 
results about the existence and uniqueness of a surface wave. 
For a general isotropic constraint \eqref{Constraint}, 
$\delta^* / \gamma^*$ is not necessarily equal to 1, but his method, 
together with the S-E inequalities (\ref{SE2}), can nevertheless be 
directly transposed to our problem.
In summary, we obtain the following results.

The {\em limiting speed} $\hat{v}$, which is the upper bound of the 
subsonic interval $I=[0, \hat{v}]$ for $v$ where the decay requirement 
$\Re(s_i) >0$ is satisfied, is defined by
\begin{equation}
\overline{\rho} \hat{v}^2 = 
\left\{
\begin{array}{lll}
& \alpha^*,  & \rm {when} \;\;\; 2\beta^* > \alpha^*, 
 \\
&  2[\beta^*-\gamma^* + \sqrt{\gamma^*(\alpha^*-2\beta^*+\gamma^*)}]
 < \alpha^*, & \rm {when} \;\;\;  2\beta^* < \alpha^*.
\end{array}
\right.
\end{equation}

A necessary and sufficient condition of existence for a root $v_R$ in 
$I$ is
\begin{equation} \label{conditions}
f(0)>0, \;\;\; f(\hat{v})<0.
\end{equation}

Finally, when a root exists, it is unique.

\section{The bifurcation criterion in special cases}
\setcounter{equation}{0}

We specialize the bifurcation criterion (\ref{bifurcation}) to each of 
the four examples of isotropic constraints considered in this paper 
and pick special forms of the strain energy in order to obtain 
explicit results.  
In each case we find that the bifurcation criterion is of a 
(relative) universal nature in that it depends only on the principal 
stretches, not on the material constants.

 \subsubsection*{Incompressibility (I)}  
For incompressibility,  the bifurcation criterion 
(\ref{bifurcation})  reduces to  \cite{DoOg90}:
\begin{equation} \label{4.1}
 \lambda_1^2W_{11} - 2\lambda_1\lambda_2W_{12}
  + \lambda_2^2 W_{22} +\lambda_2[W_1+(2-\lambda_1^{-1}\lambda_2)W_2]
 = 0.
\end{equation}

We consider the strain energy function for ``generalized Varga 
materials", 
\begin{equation} 
W=d_1(\lambda_1 +\lambda_2 +\lambda_3 - 3) 
+ d_2(\lambda_1\lambda_2 +\lambda_2\lambda_3 +\lambda_3\lambda_1 -3),
\label{4.2}
\end{equation}
in which $d_1$ and $d_2$ are material constants.  
The S-E conditions \eqref{SE2} lead to $d_1 + 2d_2 \lambda_3 >0$, and 
so $d_1>0$, $d_2 \ge 0$ or $d_1 \ge 0$, $d_2 > 0$. 
The bifurcation criterion (\ref{4.1}) leads to
\begin{equation}
(d_1+\lambda_3d_2)(3\lambda_1-\lambda_2)=0,
\quad \text{or} \quad 
\lambda_2=3\lambda_1,
\label{4.5}
\end{equation}
independently of the choice of material constants.

\subsubsection*{Bell's constraint (II)}  
For a Bell constrained material, (\ref{4.1}) reduces to the 
bifurcation criterion 
\cite{Dest03}:
\begin{equation} \label{4.9}
\lambda_1(W_{11} - 2 W_{12} +  W_{22}) + W_1 - W_2 = 0.
\end{equation}
For the specific example of the ``simple hyperelastic Bell material'' 
\cite{Beat01} we take 
\begin{equation} 
W=d_2(\lambda_1\lambda_2 + \lambda_2\lambda_3 + \lambda_3\lambda_1 - 3)
+ d_3(\lambda_1\lambda_2\lambda_3-1).
\label{4.10}
\end{equation}
Here the S-E conditions \eqref{SE2} lead to 
$-(d_2 + 2d_3 \lambda_3) >0$, and 
so $d_2<0$, $d_3 \le 0$ or $d_2 \le 0$, $d_3 < 0$,  
while the bifurcation criterion (\ref{4.1}) reduces to
\begin{equation}
(d_2+\lambda_3d_3)(\lambda_2 -3\lambda_1)=0,
\end{equation}
leading again to (\ref{4.5}).

\subsubsection*{Areal constraint (III)}  
The bifurcation criterion (\ref{4.1}) reduces to
\begin{multline} \label{areal} 
(\lambda_1+\lambda_3)^2W_{11}
 - 2(\lambda_1+\lambda_3)(\lambda_2+\lambda_3)W_{12} + 
      (\lambda_2+\lambda_3)^2W_{22}
\\ 
  + \lambda^{-1}_1 (\lambda_1+\lambda_3)(\lambda_2+\lambda_3)W_1
   +  (\lambda_2+\lambda_3)[2 -\lambda^{-1}_1(\lambda_2+\lambda_3)]W_2 
     =0.
\end{multline}
For the following material, which we term ``simple hyperelastic areal 
material'',
\begin{equation} \label{simpleAreal}
W=d_1(\lambda_1 +\lambda_2 +\lambda_3 - 3) 
+ d_3(\lambda_1\lambda_2\lambda_3  - 1),
\end{equation}
the S-E conditions \eqref{SE2} lead to $d_1 - d_3 \lambda_3 >0$, and 
so $d_1>0$, $d_3 \le 0$ or $d_1 \ge 0$, $d_3 < 0$.  
The bifurcation criterion (\ref{areal}) reduces to
\begin{equation} 
(d_1-\lambda^2_3d_3)(3\lambda_1-\lambda_2)=0,
\end{equation}
leading once more to (\ref{4.5}).

\subsubsection*{Ericksen's constraint (IV)}  
For Ericksen materials the bifurcation criterion (\ref{4.1}) reduces 
to
\begin{equation} \label{Ericksen}
 \lambda_2^2 W_{11} - 2\lambda_1\lambda_2 W_{12}+\lambda_1^2 W_{22} 
 +\lambda_2[W_1 - \lambda_2^{-2}(\lambda_1^ 2 + \lambda_1\lambda_2 
      + \lambda_2^2)W_2]
 = 0.
\end{equation}
For the following ``simple hyperelastic Ericksen material'',
\begin{equation} \label{simpleEricksen}
 W=
 D_2(\lambda^2_1\lambda^2_2 + \lambda^2_2\lambda^2_3 
      + \lambda^2_3\lambda^2_1 - 3)
 + D_3 (\lambda^2_1\lambda^2_2\lambda^2_3  - 1),
\end{equation}
the S-E conditions \eqref{SE2} lead to $-(D_2 + D_3 \lambda_3^2) >0$, 
and so $D_2 < 0$, $D_3 \le 0$ or $D_2 \le 0$, $D_3 > 0$.  
The bifurcation criterion (\ref{Ericksen}) simplifies to
\begin{equation}\label{4.16}
\lambda_1^3 + 5 \lambda_1^2 \lambda_2 
 -  \lambda_1\lambda_2^2 - \lambda_2^2 = 0.
\end{equation}

\subsubsection*{Summary}
For the four specific constraints and strain-energy functions above, 
we found relative-universal bifurcation criteria, even though two 
independent material constants were involved. 
Moreover, the bifurcation criterion for each of 
the first three forms of $W$ turns out to be 
the same, although the constraints are different; 
of course, the corresponding critical stretch ratios differ in each 
case because they are obtained by solving the bifurcation criterion in 
conjunction with the constraint condition. 
These differences are highlighted in Table 1, where the critical 
stretch ratios $(\lambda_1)_{\text{cr}}$ for surface stability in 
compression of the four constrained materials presented in this 
Section have been computed in the case of plane strain $\lambda_3=1$, 
and of equibiaxial strains 
$\lambda_2=\lambda_3$ and   $\lambda_1=\lambda_3$.
All in all, it seems that simple hyperelastic Bell materials can be 
compressed the most before the bifurcation criterion is reached, 
whilst Ericksen materials of type \eqref{simpleEricksen} can be 
compressed the least.

\begin{center}
Table 1: Critical stretch ratios $(\lambda_1)_{\text{cr}}$ for surface 
stability
\\
\vspace{5pt}
{\normalsize
\noindent
\begin{tabular}{l c c c}
\hline\rule[-3mm]{0mm}{8mm}
  & $\lambda_2=\lambda_3$   & $\lambda_3=1$   & $\lambda_1=\lambda_3$
\\
\hline
simple Bell                 & 0.429 & 0.500 & 0.600 \\
simple areal               & 0.447 & 0.535 & 0.655 \\
generalized Varga      & 0.481 & 0.577 & 0.693 \\  
simple Ericksen         & 0.603 & 0.658 & 0.730 \\ \hline
\end{tabular}
}
\end{center}

Some questions worth investigating
are raised by the previous results:  what is the largest class of 
energy function for which the result is universal for any given 
constraint? 
for what energy functions is the bifurcation criterion the same for 
two (or more) different constraints (separately or together)? 
for a material whose energy function is  linear in the invariants of 
the stretch tensor such as \eqref{4.2}, 
\eqref{4.10}, \eqref{simpleAreal}, which constraints lead to the 
same bifurcation criterion \eqref{4.5}? and so on.
Unfortunately we must, because of space limitations, leave these 
problems open.

\subsubsection*{Concluding remark: an incompressible model 
for human thoracic aorta}

We conclude with an example of a non-universal bifurcation criterion 
using a model taken from the biomechanics literature.
Horgan and Saccomandi \cite{HoSa03} recently discussed the merits 
of the ``limiting chain extensibility model'' proposed by Gent 
\cite{Gent96} and its applicability to the modelling of 
strain-stiffening biological tissues. 
For this incompressible (I) material, the strain energy is 
\begin{equation} \label{Gent} 
W = \textstyle{\frac{1}{2}} 
  \mu J_m 
   \ln \left(1 - 
    \dfrac{\lambda_1^2 + \lambda_2^2 + \lambda_3^2 - 3}{J_m} \right), 
\quad 
\lambda_1^2 + \lambda_2^2 + \lambda_3^2 < 3 + J_m,
\end{equation}
where $\mu$ is the shear modulus and $J_m$ is a constant. 
The smaller $J_m$ is, the stiffer the material becomes; 
conversely the body becomes more deformable as $J_m$ 
increases, with the neo-Hookean model as a limit for 
$J_m \rightarrow \infty$, where there are no restrictions on the 
values that the stretch ratios may take. 

For the Gent model \eqref{Gent} the S-E conditions \eqref{SE2} lead to 
$\mu J_m >0$ so that $\mu$ and $J_m$ are of the same sign.  
The bifurcation criterion (\ref{4.1}) yields
\begin{equation} \label{bifurcGent}
f(\lambda_1, \lambda_2)(J_m + 3 -  \lambda_3^2)  
 -  f(\lambda_2, \lambda_1)
       (\lambda_1^2 + 2\lambda_1\lambda_2 - \lambda_2^2) = 0,
\end{equation} 
where $f(x,y) := x^3 + x^2 y + 3 xy^2 - y^3$. 
Note that as $J_m \rightarrow \infty$ this equality tends to 
$f(\lambda_1, \lambda_2) =0$, which is the universal bifurcation 
criterion for neo-Hookean (and Mooney-Rivlin) materials \cite{Biot65}. 
Using experimental data for the thoracic aorta of a 21-year old male 
and of a 70-year old male, Horgan and Saccomandi \cite{HoSa03} 
found that for these samples, $J_m = 2.289, 0.422$, respectively. 
For these values of $J_m$ and for the three special prestrains 
where  $\lambda_3=1$, or $\lambda_2=\lambda_3$, or $\lambda_1=\lambda_3$, 
we find that there exists no value of $\lambda_1$ such that the 
bifurcation criterion is satisfied, indicating that the aorta of the 
two males is \textit{always stable} near the surface with respect to 
finite compressions of these types. 
Keep in mind however that the range of possible ratios is limited by 
the inequality \eqref{Gent}$_2$. 
For example, in plane strain $\lambda_3=1$ (no extension in the $X_3$ 
direction), we find that 
\begin{equation} 
\sqrt{1 + J_m(1 - \sqrt{1+4J_m^{-1}})/2}
   < \lambda_{1,2} 
    <  \sqrt{1 + J_m(1 + \sqrt{1+4J_m^{-1}})/2}. 
\end{equation}  
For the younger aorta, this range is [0.497,2.010]; 
for the older, stiffer, aorta, the range is [0.727,1.376].


\end{document}